\documentclass[conference]{IEEEtran}

\usepackage{hyperref}
\usepackage{cite}
\usepackage{grffile}
\usepackage{amsmath}
\usepackage{amsfonts}
\usepackage{amssymb}
\usepackage{amssymb,amsmath,amsfonts,lineno,setspace, MnSymbol}
\usepackage{epsfig,graphicx,multirow,graphics,ifpdf,subfigure,subfloat,float}
\usepackage{graphics} 
\usepackage{wrapfig}
\usepackage[usenames,dvipsnames]{xcolor}
\usepackage{url}

\renewcommand\IEEEkeywordsname{Keywords}

\pdfoutput=1

\usepackage{ifxetex}
\ifxetex
  \usepackage[protrusion]{microtype}
\else
  \usepackage[expansion,protrusion]{microtype}
\fi

\begin{document}
\title{Scalable Architecture for Anomaly Detection\\ and Visualization in Power Generating Assets}

\author{
  \IEEEauthorblockN{
    Paras Jain\IEEEauthorrefmark{1},
    Chirag Tailor\IEEEauthorrefmark{1},
    Sam Ford\IEEEauthorrefmark{1},
    Liexiao (Richard) Ding\IEEEauthorrefmark{3},
    Michael Phillips\IEEEauthorrefmark{3},\\
    Fang (Cherry) Liu\IEEEauthorrefmark{4},
    Nagi Gebraeel\IEEEauthorrefmark{3},
    Duen Horng (Polo) Chau\IEEEauthorrefmark{1}
  }
  \IEEEauthorblockA{
    \IEEEauthorrefmark{1}College of Computing\\
  }
  \IEEEauthorblockA{
    \IEEEauthorrefmark{3}H. Milton Stewart School of Industrial \& Systems Engineering\\
  }
  \IEEEauthorblockA{
    \IEEEauthorrefmark{4}Partnership for an Advanced Computing Environment\\
  }
  Georgia Institute of Technology\\
  Atlanta, Georgia, USA.\\
  Email: \{paras, chirag.tailor, sford100, richard.ding, mphillips68, polo\}@gatech.edu,\\ fang.liu@oit.gatech.edu, nagi.gebraeel@isye.gatech.edu
}

\maketitle

\begin{abstract}
Power-generating assets (e.g., jet engines, gas turbines) are often instrumented with tens to hundreds of sensors for monitoring physical and performance degradation.  Anomaly detection algorithms highlight deviations from predetermined benchmarks with the goal of detecting incipient faults.

We are developing an  integrated  system  to  address  three key challenges within  analyzing sensor data  from  power-generating assets:
(1) difficulty in ingesting and analyzing data from large numbers of machines;
(2) prevalence  of  false  alarms  generated  by  anomaly  detection algorithms resulting in unnecessary downtime and maintenance; and 
(3) lack of an integrated visualization that helps users understand and explore the flagged anomalies and relevant sensor context in the energy domain.

We present preliminary results and our key findings in addressing these challenges.
Our system's scalable event ingestion framework, based on OpenTSDB, ingests nearly 400,000 sensor data samples per seconds using a 30 machine cluster.
To reduce false alarm rates, we leverage the False Discovery Rate (FDR) algorithm which significantly reduces the number of false alarms.
Our visualization tool presents the anomalies and associated content flagged by the FDR algorithm to inform users and practitioners in their decision making process.  

We believe our integrated platform will help reduce maintenance costs significantly while increasing asset lifespan. 
We are working to extend our system on multiple fronts, such as scaling to more data and more compute nodes (70 in total).


\end{abstract}


\begin{figure*}[t!]
\centering
\includegraphics[width=0.7\textwidth]{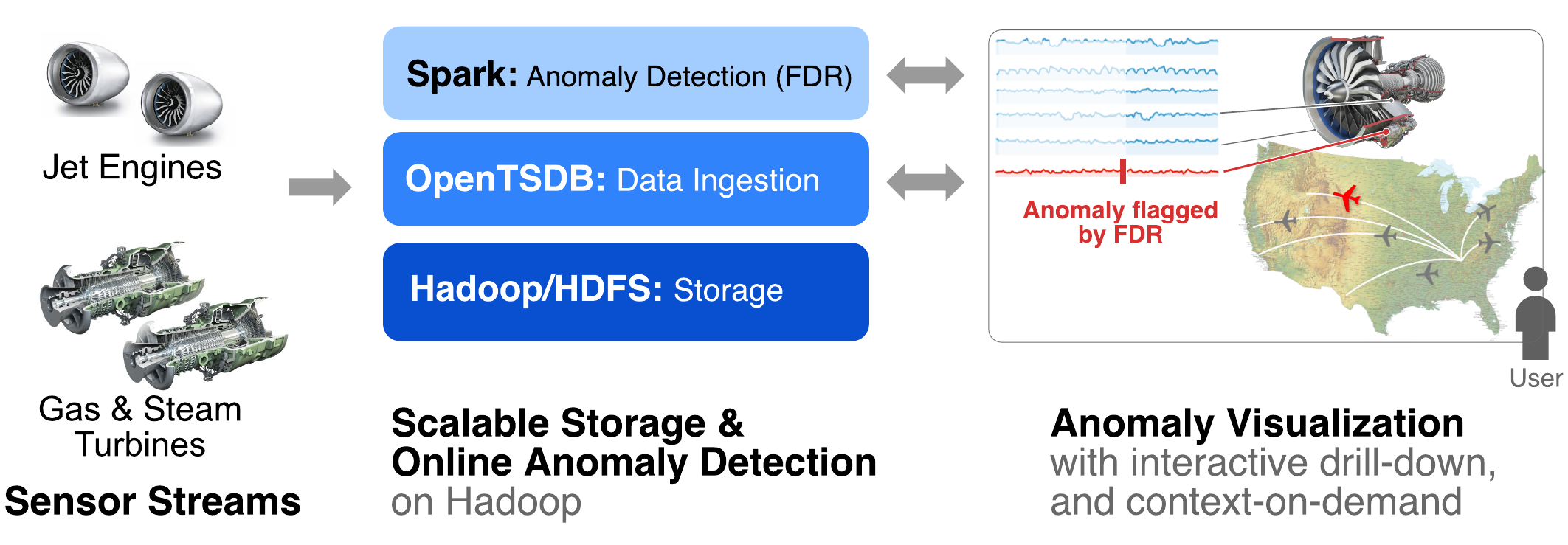}
\caption{Overview of our system architecture, with a scalable Hadoop-based storage and anomaly detection backend, and an interactive visual frontend to aid user understanding.} 
\label{fig:overview}
\end{figure*}

\section{Introduction}


To improve public safety, modern power generating assets (e.g., jet engines, gas turbines) are instrumented with hundreds of sensors to monitor  physical performance degradation.  Such sensors, such as temperature or pressure, is installed with the goal of measuring potential signals of asset failure.  Given the scale of data produced, it is impossible for humans to directly monitor every signal.  Instead, this monitoring process can be automated by studying possible deviations from pre-specified benchmarks, with the goal of detecting incipient faults.

Anomaly detection involves defining a pattern in observations that represent normal behavior and declaring observations that do not belong to that region as anomalies. A multitude of detection algorithms and techniques have been developed and commercialized over the years, many of which have been applied in the manufacturing domain~\cite{fault_detection_book1, fault_detection_book2} for what has become known as Statistical Process Control (SPC).

In large scale settings involving thousands of power generating assets where each asset is monitored by a large number of sensors, the problem of false alarms becomes a significant challenge.  
False alarms can be very costly --- for example, 50\% of replaced parts in aircraft are classified as ``no fault found''~\cite{no_fault_found}. In energy and aerospace domains, reducing rates of false alarms can measurably reduce lifetime maintenance costs for power generating assets.




Our goal is to develop an integrated system
that reduces false alarms in multi-stream condition monitoring of power generating assets
using a scalable analytics architecture that ingests, stores and analyzes large amounts of sensor data, and interactively visualize the computation results to enhance user understanding and advance decision making capabilities. 

Our ongoing work's contributions are:
\begin{itemize}
\item We present a scalable event ingestion and storage architecture that can handle 399,000 sensor samples per second with a 30 node storage cluster.
\item We adapt the FDR algorithm to the energy domain in order to reduce the rate of falsely identified anomalies.
\item We demonstrate a visualization tool that enables interactive exploration of power-generating asset sensor data with associated anomalies.
\end{itemize}

\section{System Overview}
Our proposed architecture consists of three key components: 
the anomaly detection algorithm, streaming sensor data ingestion, 
and interactive visualization. \autoref{fig:overview} summarizes our envisioned system architecture.

Data storage is non-trivial with large scale deployments of assets generating huge volumes of data. Our system will need to eventually scale to process hundreds of thousands of sensor samples per second. 
After evaluating candidate storage solutions, we chose to build our platform over the Open Time Series Database (OpenTSDB)~\cite{web:opentsdb} to store both streaming sensor data as well as flagged anomalies (Section~\ref{sec:opentsdb}). 
OpenTSDB is supported by HBase~\cite{web:hbase} as the underlying distributed file storage system.

The choice of anomaly detection algorithm is central to the project. A survey~\cite{chandola2009anomaly} of techniques for anomaly detection divides traditional algorithms into several categories: rule-based systems, statistical techniques, spectral techniques and neural networks. As the primary criteria for anomaly detection is to control the rate of false positives, we specifically chose the False Discovery Rate algorithm~\cite{FDRBenjamini_1995, FDRBenjamini_2001} (described in~\autoref{sec:anomalydetection}). 
Offline evaluation of the anomaly detection algorithm currently executes in the Spark framework \cite{zaharia2010spark} in batch mode.
Given its rich distributed matrix computation libraries, 
Spark is a natural choice for evaluating the FDR algorithm.

Both the online anomaly detection and streaming sensor data ingestion components run on the Data Science Platform (DSP)~\cite{DSPLiu_2016} built at Georgia Tech College of Computing. The online anamoly detection component runs on a 44 nodes HDFS/Hadoop/Spark cluster, while the streaming sensor data ingestion runs on a 32 nodes HDFS/HBase/OpenTSDB cluster.

Our visualization tool allows for operators to interactively explore large volumes of time-series data (\autoref{sec:vis}). By enriching sensor data with potential anomalies and integrated real-time analytics, the platform can serve as a powerful control center to monitor faults across large sensor networks.

\subsection{Evaluation Dataset}
\label{sec:datasets}
Real datasets from industry partners contain actual sensor readings from power generating assets like gas turbine and jet engines.
However, these datasets often include sensitive information, and currently are not available for off-site evaluation.
As our very first goals of developing the system are to investigate the algorithmic scalability, visualization capabilities, and hardware platform requirements, access to a real dataset is not critical to our current stage of investigation.
Therefore, we generated a dataset for training and evaluation of the algorithm. 
This allows measuring the exact degree to which FDR reduces false alarm rates while allowing us to verify the algorithms ability to detect various classes of injected faults.

The training dataset contains 100 simulated units, each with 1000 sensors (on order with the 3000 sensors in the Siemens SGT5-8000H gas turbine~\cite{ratliff2007new}). We modeled three primary categories of faults:
\begin{itemize}
\item Pure random noise for comparison
\item Pure random noise plus gradual degradation signal
\item Pure random noise plus sharp shift
\end{itemize}

Injected faults are correlated across sensors which allows measuring the algorithm's response to deviations across multiple signals.

\section{Scalable Data Ingestion \& Storage}
\label{sec:opentsdb}

Processing and storing sensor samples is non-trivial at the scale of a production deployment.
We expect our system will need to ingest and analyze at least 100,000 sensor samples per second (based on estimation in~\autoref{sec:datasets}, assuming each sensor will generate data at 1Hz).

This project utilizes OpenTSDB~\cite{web:opentsdb}, an open-source, scalable, time series database which leverages HBase~\cite{web:hbase}, an Apache top level project inspired by Google's BigTable~\cite{bigtable}, to manage data in a distributed manner and provide horizontal scalability. 
We chose OpenTSDB because
it allows us to easily horizontally scale out our system to more storage nodes, while maintaining a stable, linear scaleup in streaming ingestion  (as we shall describe in Section~\ref{sec:ingestionresults}). 



\begin{figure*}[t]
\centering
\includegraphics[width=\textwidth]{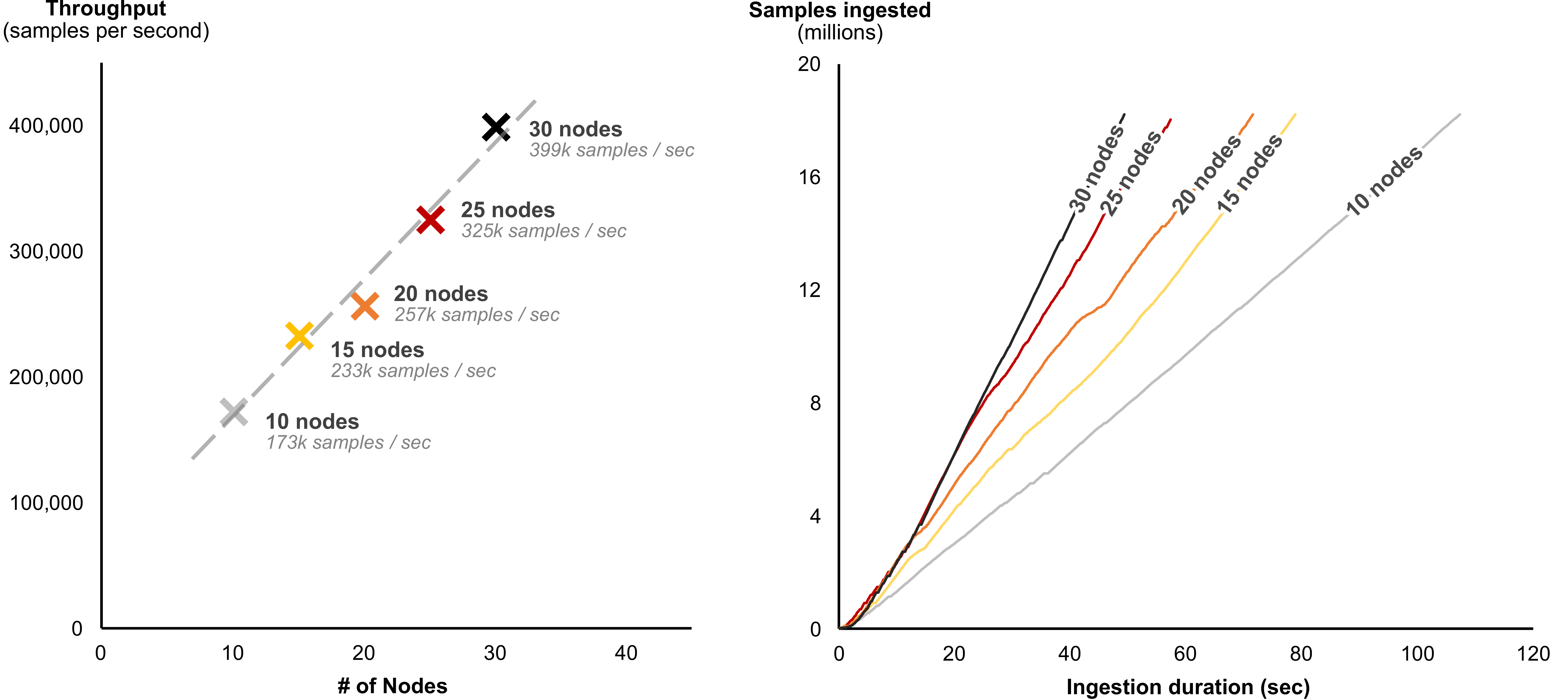}
\caption{\textbf{Left:} Ingestion rates achieved as the ingestion framework was ramped up from 10 machines to 30 machines. The system scales linearly, with each added machine increasing throughput by 11K samples per second on average.
\textbf{Right:} The line graph of sensor samples ingested versus the ingestion duration shows a constant and stable ingestion rate for each configuration of the framework.}
\label{fig:ingestiongraph}
\end{figure*}

\begin{figure*}[bt]
\centering
\includegraphics[width=\textwidth]{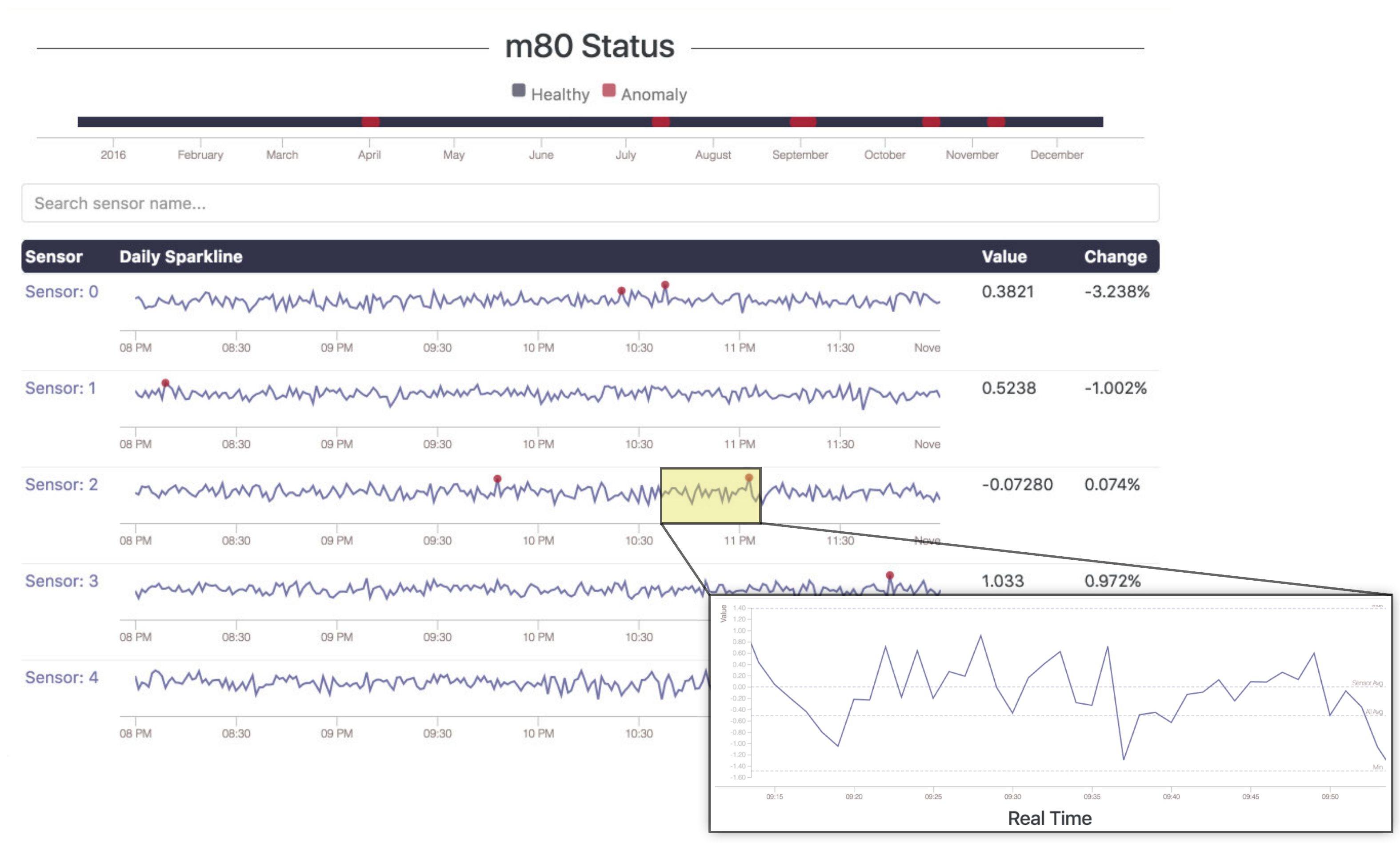}
\caption{An overview of the machine page showing sample sensor readings for machine 80. The time line of values show real time values for each sensor of the machine and points where anomalies occurred are flagged in red.}
\label{fig:singlemachine}
\end{figure*}

\subsection{Streaming Event Ingestion and Storage using OpenTSDB}

For the purposes of this preliminary work, a distributed system of 32 nodes was deployed running HDFS and HBase. HDFS was set up with one NameNode (co-running Hbase master), one Secondary NameNode, one Hbase backup master and 29 DataNodes. HBase is configured with one HMaster, one BackupHMaster, and 29 Regionservers that communicate through the built-in Apache Zookeeper~\cite{hunt2010zookeeper} coordination service. Each node is also running an instance of a TSD Daemon for time series data writing and querying. 

OpenTSDB organizes time series data into metrics and allows for the assignment of multiple tags per metric. The tags provide unique identifiers for querying data and allow the data to be compartmentalized into sub series. The simulated data generated for this project is stored into a metric called ``energy'' with tags for ``unit'' and ``sensor''. 

For storing data, the TSD Daemon takes a metric, timestamp, data value, and tag identifiers as input and produces an entry to be written to an HBase table. First, the key is generated from a binary encoding of the metric, timestamp, and tag values. Then, the TSD Daemon submits an RPC call to HBase which distributes the write based upon the key value. The writes for similar keys are grouped onto the same Regionserver by HBase.

\subsection{Preliminary Results \& Key Findings}
\label{sec:ingestionresults}
\noindent \textbf{Linear Scale-up.} We tested our event ingestion framework with the evaluation dataset (described in Section~\ref{sec:datasets}, consisting of 100 assets with 1000 sensors each). \autoref{fig:ingestiongraph}a shows the ingestion rate scales up linearly with the number of machines. Each machine runs one HBase RegionServer instance and one OpenTSDB daemon. 
The ingestion rate reaches 300,000 samples per second with 30 machines, while maintaining a stable ingestion speed, shown in~\autoref{fig:ingestiongraph}b.

\noindent\textbf{OpenTSDB Key Design.} 
One obstacle encountered early in the ingestion process was an issue with these writes not being distributed across all the HBase Regionservers efficiently. Since sequential data values share the same metric and similar timestamp values, their binary encoded keys are also similar resulting in the RPC calls being sent to the same HBase Regionserver. To combat this, the binary key encodings were salted with an additional uniformly randomly generated byte at the beginning to create unique keys for chronological data values. Additionally, HBase regions were manually split to ensure each region handled an equal proportion of the writes. Salting the keys allowed for the full utilization of all the deployed HBase Regionservers and provided a dramatic increase to the ingestion rate.

\noindent \textbf{Buffering Requests for Backpressure and Scalability.} 
Another obstacle encountered while working with OpenTSDB and HBase was frequent crashes of Regionservers due to overloaded RPC Queues. 
Initially, the cause of these crashes was attributed to HBase having no means of providing back pressure to RPC calls made by OpenTSDB. 
To remedy this, we built a reverse proxy to buffer requests to OpenTSDB in order to limit the number of concurrent requests. Compaction was also disabled on OpenTSDB to reduce RPC calls to HBase.
This proxy also serves to increase ingestion throughput by load-balancing traffic to multiple ingestion processes.  Ingestion throughput scales horizontally by distributing the requests to the OpenTSDB nodes via a round-robin fashion.

\section{Flagging Anomalies with\\ Low False Alarm Rates}
\label{sec:anomalydetection}

A key component in our system is the algorithm to flag potential anomalies. Given the large expense caused by erroneously flagged faults, we aim to use an anomaly detection algorithm that balances identifying the majority of true faults while also controlling the rate of false alarms.

From a statistical standpoint, anomaly detection amounts to performing a hypothesis test on sample observations to detect possible shifts in the mean of the sampling distribution.  Rejection of the null hypothesis implies that there is significant evidence to conclude that the distribution has indeed changed. A common mistake committed in hypothesis testing is to reject the null hypothesis when it is actually true (type I error).

In our setting, a type I error amounts to a false alarm, i.e., an equipment is classified as faulty when in reality there was nothing wrong.  One of the key aspects of type I errors is that they tend to increase as the number of hypothesis tests increases. In our context, this translates to higher false alarm rates as the number of sensors increases. For example, for a single sensor with an allowable $\alpha=0.05$, the  probability of making at least one false alarm is $5\%$. However, if we increase the number of sensors to $10$ sensors each with $\alpha=0.05$, that probability  jumps to $40\%$ , i.e., $1-(1-\alpha)^{10} = 0.4$.

Traditionally, false alarms in multi-inference studies were controlled using family-wise error rate (FWER).  FWER focuses on controlling the probability of committing any type I error when performing multiple hypothesis tests by applying a correction to a family of inferences.  A popular example is the Bonferroni correction~\cite{dunn1961fwer} where for $m$ hypotheses each having a probability $\alpha$ of committing type I error, then the corrected probability for the family would be $\alpha/m$.  In other words, reject all hypotheses with \emph{p}-values $\leq \alpha/m$. One drawback of this approach was that it provided much less detection power and was overly conservative.

FDR was first introduced by Benjamini and Hochberg in 1995~\cite{FDRBenjamini_1995, FDRBenjamini_2001}.  The goal was to reduce false alarms in multiple inferences (hypothesis tests) in clinical trials. Compared to FWER, FDR was designed to control the expected proportion of type I errors.

The underlying premise of FDR is that when multiple tested hypotheses are rejected, it is more important to control the proportion of errors (wrong rejections) than it is a single erroneous rejection. As we wish to balance the rate of type I and type II errors, FDR is a promising option for a choice of algorithm for preliminary testing.

\subsection{Preliminary Results}
Our implementation of the FDR algorithm is composed of two parts --- an offline training component and an online evaluation component. 
Offline training occurs in Spark, running in batch mode. 
Spark's MLlib~\cite{meng2016mllib} provides an implementation of distributed matrix factorization, which allows our offline training system to scale to large numbers of sensors. Once training is complete, we can evaluate for anomalies at a rate of 939,000 sensor samples per second on average.

In offline training, model estimation of each sensor on each unit begins by calculating the covariance matrix of each data set. Singular Value Decomposition is then performed on each covariance matrix to obtain the mean and variance. Results from the decomposition are cached to HDFS. Evaluation is thereby relatively fast requiring a single matrix multiplication per iteration. Results from online evaluation are reported back to OpenTSDB for use by the integrated visualization tool. The current system can deal with one machine at a time and we plan to utilize concurrency of Spark to scale up workload. 

\section{Anomaly Visualization}
\label{sec:vis}

Advanced anomaly detection without commensurate visualization presents limited value for operators. Our platform includes an interactive visualization tool that equips users to quickly respond to flagged anomalies. Our tool integrates A) live sensor data, B) highlighted anomalies, and C) real-time system analytics into a single control center for users to monitor and react to events in a network of power-generating assets.

\subsection{Preliminary Results}

Our tool provides an overview of the overall health of the network of power-generating assets. It helps users explore and understand the context surrounding the flagged anomalies. By using the FDR anomaly detection algorithm, we avoid unnecessarily notifying users of false alarms.

Analytics summarize global system status across a large deployment of power-generating assets. By selectively surfacing the most concerning anomalies, we allow users to focus only on what is important. Unit status is summarized neatly into a single status bar as seen at the top of~\autoref{fig:singlemachine}.

Power-generating asset faults often result in correlated anomalies across multiple sensors (e.g., pressure and temperature). Our tool displays all sensor readings with relevant anomalies annotated directly on a compact sparkline chart as seen in the center of~\autoref{fig:singlemachine}.

Drill-down capabilities enable users to quickly examine details about a fault with necessary context. Operators can click on anomalies which surfaces a detailed view of the sensor data, as shown at at the bottom of~\autoref{fig:singlemachine}.

The visualization tool is a web application that is available on both desktop and mobile devices. Mobile access allows technicians to explore pertinent sensor data while performing maintenance on a particular machine in the field.

\section{Conclusion \& Ongoing Work}

We present our preliminary work and key findings in developing 
an  integrated  system  to  address  three key challenges within  analyzing sensor data  from  power-generating assets: 
(1) difficulty in ingesting and analyzing data from large numbers of machines;
(2) prevalence  of  false  alarms  generated  by  anomaly  detection algorithms resulting in unnecessary downtime and maintenance; and 
(3) lack of an integrated visualization that helps users understand and explore the flagged anomalies and relevant sensor context in the energy domain.
%
%
Our system can currently ingest and analyze 399,000 sensor samples per second while running on a 30 node cluster. The system visualizes sensor data in an interactive web application which presents potential anomalies with the associated context surrounding the event.

Ongoing work for the project includes: 
experimenting with increasing storage nodes to further scale up throughput, 
migrating our anomaly detection implementation to Spark Streaming~\cite{zaharia2013discretized} for online training,
and evaluating our system with domain users through our collaboration with industry partners like General Electric (GE) to test the system on their datasets.

\section*{Acknowledgment}
This work is supported in part by the Strategic Energy Institute (SEI) at Georgia Tech, and NSF grants IIS-1563816, TWC-1526254, IIS-1217559.
We also thank Yahoo! for their generous 200-machine donation.
We thank Will Powell on hardware support for our system.




\bibliographystyle{IEEEtran}
\bibliography{ref}

\end{document}